\documentstyle[prb,preprint,aps]{revtex}
\begin{document}
\draft
\title{
First-principles calculation of exchange interaction 
and exchange force between magnetic Fe films
}
\author{
Kohji Nakamura, Hideo Hasegawa$^1$, Tamio Oguchi$^2$,  
Kazuhisa Sueoka$^3$, \\
Kazunobu Hayakawa, and Koichi Mukasa$^{3,4}$
}
\address{
Catalysis Research Center, Hokkaido University,
Sapporo 060, Japan \\
$^1$Department of Physics, Tokyo Gakugei University,
Koganei, Tokyo 184, Japan.\\
$^2$Department of Materials Science, Hiroshima University,
Higashi-Hiroshima 739, Japan.\\
$^3$Nanoelectronics Laboratory, Graduate School of Engineering,
Hokkaido University,
Sapporo 060, Japan. \\
$^4$CREST "Spin Investigation Team", 
Japan Science and Technology Corporation (JST).
}
\maketitle
\begin{abstract}
First-principles calculations have been performed  for the
exchange interaction and exchange force 
between two magnetic Fe(100) thin films. 
It is shown that the magnitude of the calculated exchange force
is of the order of $10^{-9} $N at $d/a < 1$ and
of the order of $10^{-10} $N at $1 < d/a < 1.5$,
where $d$ denotes the distance between surfaces
of the two thin films and $a$ the lattice constant of bulk Fe.
The obtained forces are sufficiently larger than 
the sensitivity of the current
atomic force microscopy of $10^{-12} - 10^{-13}$N, 
which suggests the feasibility of the 
{\it exchange force microscopy}.
\end{abstract}

\vspace{1.0cm}
%
%

\begin{center}
{\bf I. INTRODUCTION}
\end{center}

Surface magnetism has attracted much attention both in fundamental 
and applied researches. 
Experimental techniques to investigate the
magnetic structure of the surface having  been developed,
mainly detect the spin polarization of electrons emitted from surface and
the average magnetization on sample surface within a sub-micron range
is usually observed.
None of the techniques has so far
probed the local magnetic structure in an atomic scale. 

The scanning tunneling microscopy (STM)\cite{82Bin}
and the atomic force microscopy (AFM)\cite{86Bin} 
provide us with a way to investigate 
the surface structure with atomic resolution. 
The STM probes the surface of conducting materials 
by measuring tunneling 
current
between the tip and sample, whereas  the AFM
senses the atomic force between them. 
The AFM has a great advantage that it can be applied not only to 
conducting materials but also to insulators.
The extension of the STM to measurements of the magnetic structure, 
the so-called spin-polarized STM (SP-STM), 
is currently being developed 
by various groups with different approaches.\cite{90Wie,92Alv,93Sue}

The magnetic force microscopy (MFM)\cite{87Mar} 
detects the force arising from the long-range interaction between 
magnetic dipoles of the tip and sample. 
The typical tip-sample separations in the MFM are
of the order of more than 10nm and the spatial resolution 
is of the order of 10nm to 100nm. 
It is fairly certain that improvement of the resolution 
might be made by probing the short-range exchange force 
rather than the long-range magnetic 
dipole force.\cite{94Muk,95Muk,95Nes} 

In  previous papers,\cite{94Muk,95Muk} 
we pointed out the possibility of 
the {\it exchange force microscopy} (EFM) which
probes the short-range exchange force between the tip and sample.
Since microscopic understanding of the exchange interaction 
between the tip and sample
is of great importance and crucial in designing the EFM, 
we previously studied the exchange interaction 
between ferromagnetic tip and ferromagnetic sample
based on a one-dimensional electron gas model.\cite{94Muk,95Muk}
By using a tight-binding model,
Ness and Gautier\cite{95Nes} calculated the exchange interaction
between a ferromagnetic Fe tip and 
magnetic Cr(001) and Ni(001) surfaces.
Since these calculations inevitably
employ semi-phenomenological parameters,
it is highly desirable to make $quantitative$ evaluation of the
exchange interaction and the exchange force for more realistic systems
on the basis of
first-principles electronic structure calculations,
which is the purpose of the present paper.

The paper is organized as follows: In Sec.II we discuss 
the adopted model and calculation method.  
The calculated results are presented in Sec.III.
The final Sec.IV is devoted to conclusion and discussion.

\begin{center}
{\bf II. METHOD OF CALCULATION}
\end{center}

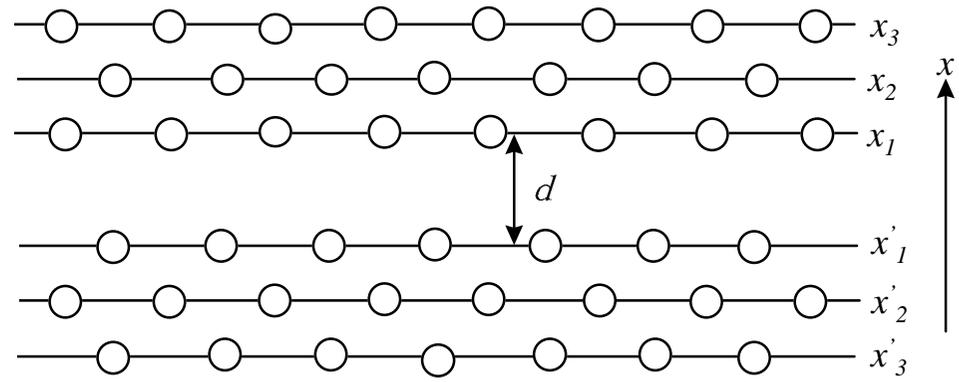
\begin{figure}
\caption{
Schematic representation of the two three-atom-layer 
Fe (100) films adopted in our calculation, 
surface atoms of the one film  facing the hollow sites 
on the surface of the other film.
Open circles represent Fe atoms 
and the lines denote the layer planes which are
referred to as the $x_n$ and $x'_n$ layers ($n = 1-3$).
}
\label{fig1}
\end{figure}

By modeling the tip and samples to be used in  actual experiments,
we adopt the two three-atomic-layer Fe(100)  films
which are separated by the distance $d$, as shown in Fig.~\ref{fig1}.
Surface atoms of the one film are assumed to be facing 
the hollow sites on the surface of the other film.
The translational symmetry is preserved  in the surface-normal ($x$) direction,
in which a set of the two three-layer films are periodically located
with the five-atomic-layer vacuum gap (repeated slab model). 
For a calculation of the electronic structure of the adopted films,
we employ the local-spin-density approximation 
to the density-functional theory, 
and the linear augmented-plane-wave (LAPW) method\cite{75And} 
with Hedin-Lundquist exchange correlation.\cite{71Hed} 
The potential and charge density are expanded 
with plane waves in the whole space 
and with spherical waves inside muffin-tin spheres, 
and the force is calculated by following the prescription given 
by Soler and Williams.\cite{89Sol} 
Because bulk Fe is in the ferromagnetic ground state,
magnetic moments in each three-layer film is taken to be in
the ferromagnetic alignment.
As far as the relative orientation of moments in the two films
are concerned, however, we consider the parallel (P) and 
anti-parallel (AP) configurations 
in order to calculate  the exchange interaction 
and the exchange force between the two magnetic films. 
Our LAPW calculations
are carried out by changing the film-film separation $d$
from 1.4\AA to 5.0\AA with an assumption that
the internal atomic coordinates of each film are rigidly fixed.

\begin{center}
{\bf III. CALCULATED RESULTS}
\end{center}

\noindent
{\it A. Magnetic Moments }

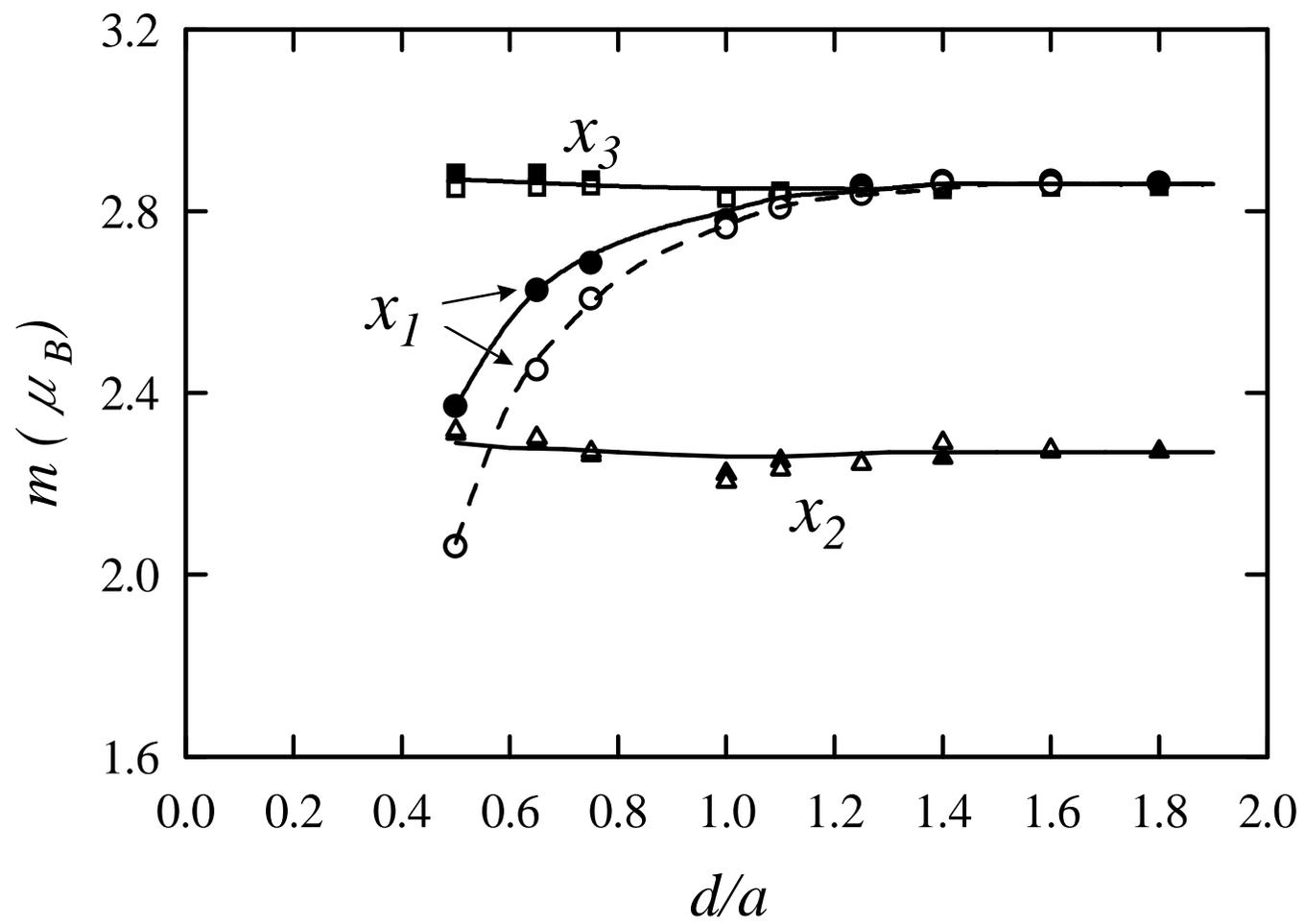
\begin{figure}
\caption{
Magnetic moments  in the muffin-tin sphere 
of an Fe atom in  the
$x_1$ (circles), $x_2$ (triangles) and $x_3$ (squares) layers 
as a function of the film-film separation $d$
normalized by the lattice constant of bulk Fe ($a$ = 2.83\AA).
Solid and open marks stand for the P 
and the AP configurations, respectively.
}
\label{fig2}
\end{figure}

We discuss first the magnetic moments in the adopted
Fe thin films.
Figure \ref{fig2} shows the spin magnetic moments calculated 
within the muffin-tin spheres 
of Fe atoms in the three layers of the upper film,
which are referred to
as the $x_1$, $x_2$ and $x_3$ layers (Fig.~\ref{fig1}).
The film-film separation $d$ in the abscissa
is denoted relative to the lattice constant of bulk Fe ($a$ = 2.83\AA).
We should note that the separation of $d/a=0.5$ 
corresponds just to the interlayer distance of bulk Fe. 
Due to the symmetry in our model,
the magnetic moments in the $x'_n$ layer of the lower film are
the same as those in the $x_n$ layer   
of the upper film in the P configuration,
while in the AP configuration those in the $x'_n$
layer has the same magnitude but with the opposite sign
as those in the $x_n$ layer.
For both P and AP configurations,
the magnetic moments at the outer $x_3$ layer
is enhanced to be about $2.9 \mu_B$,
which is almost identical to the surface magnetic moment  obtained by
slab calculations.\cite{81Wan,83Ohn}
The magnetic moments at the central $x_2$ layer is about $2.3 \mu_B$,
which is close to the experimental bulk value
and to calculated one at the central layer 
of the slab.\cite{83Ohn} 
The magnetic moments of the $x_2$ and $x_3$ layers 
for both P and AP configurations 
are almost independent of the film-film separation.
On the contrary,  magnetic moments on
the $x_1$ layer change considerably at $d<a$, 
where the magnetic moments reduce significantly 
from the surface value to the bulk one as decreasing $d$.
Nevertheless, the change in the magnitude of moments of the $x_1$ layer
becomes insignificant for $d/a > 1$.
We notice that the magnetic moments 
for the AP configuration 
are always smaller than that for the P configuration. 

\vspace{0.5cm}
\noindent
{\it B. Exchange Interaction}

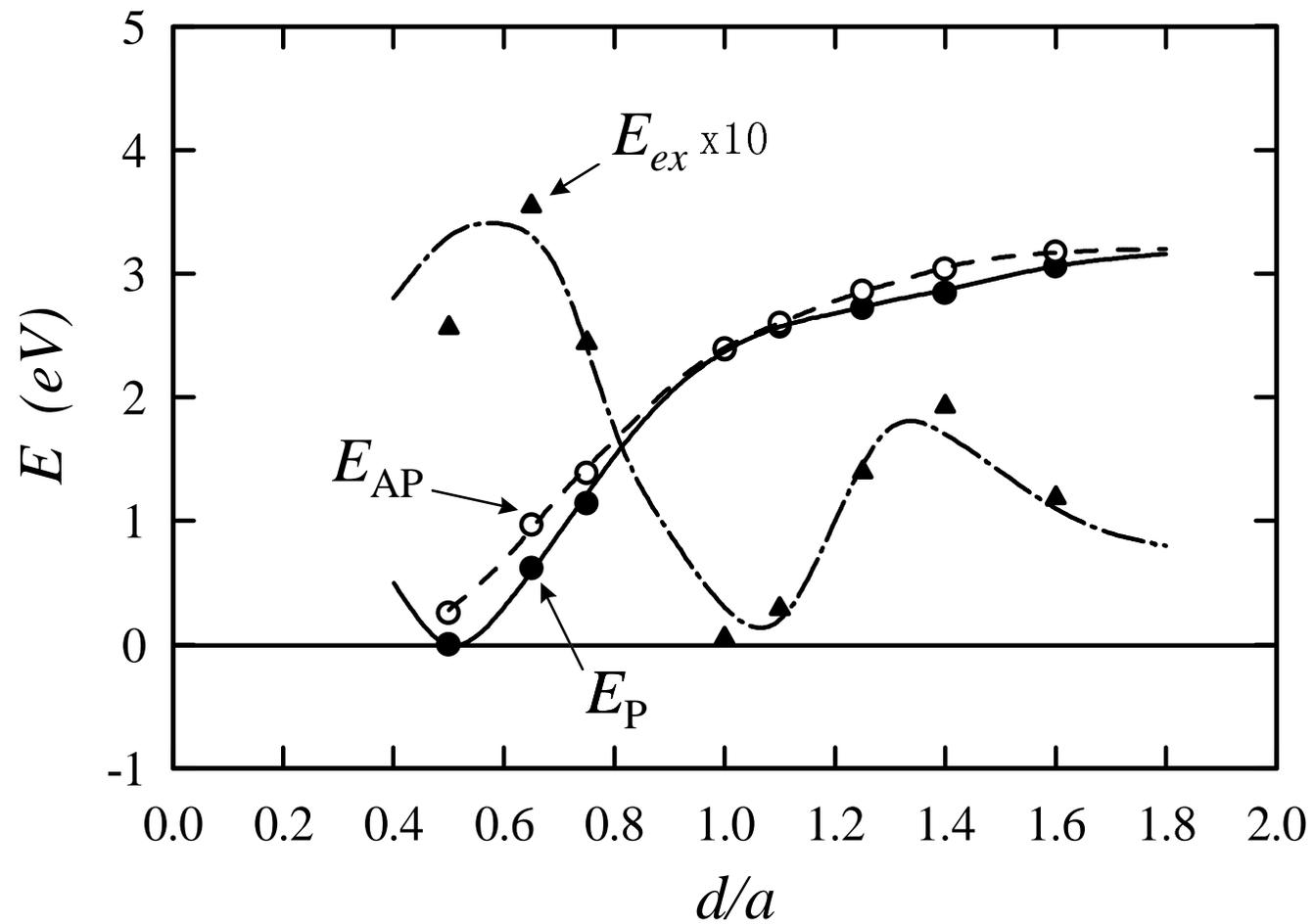
\begin{figure}
\caption{
The film-film separation dependence of
total energies in the P ($E_{\rm P}$: filled circles) 
and AF ($E_{\rm AF}$: open circles)
configurations, and of
the exchange energy defined by 
$E_{\rm ex} = E_{\rm AF} - E_{\rm P}$ (triangles).
The reference energy for $E_{\rm P}$ and $E_{\rm AF}$ is 
the total energy for the P configuration at $d/a=0.5$.
}
\label{fig3}
\end{figure}

The total energies, $E_{\rm P}$ and $E_{\rm AP}$,
for the P and AP configurations are shown in Fig.~\ref{fig3}
as a function of $d/a$. 
When the films are moved from the bulk position at $d/a = 0.5$,
the total energies for both $E_{\rm P}$ and $E_{\rm AP}$ increase.
The exchange interaction energy, $E_{\rm ex}$,
defined by $E_{\rm ex} = E_{\rm AP}-E_{\rm P}$ 
is also plotted in Fig.~\ref{fig3}. 
In the all region ($0.5 < d/a < 1.7$) investigated, 
we get the positive $E_{\rm ex}$, which means that  
the P configuration is more favorable
than the AP one.
This is expected from the fact that 
the resulting magnetic configuration corresponds to 
a natural stacking of ferromagnetic Fe in bulk.
The exchange interaction energy 
has an RKKY-type oscillation\cite{54Rud} 
with a period of about $d/a=0.7$ ($\sim$ 2\AA). 
The peak values of $E_{ex}$ are $0.35$eV at $d/a = 0.6$ and 
$0.19$eV at $d/a = 1.35$.
Note that the exchange interaction can be realized  
even at relatively far separation  
at $1.4 < d/a < 1.8$ ( 4.0\AA$<d<$5.0\AA).

\vspace{0.5cm}
\noindent
{\it C. Exchange Force}

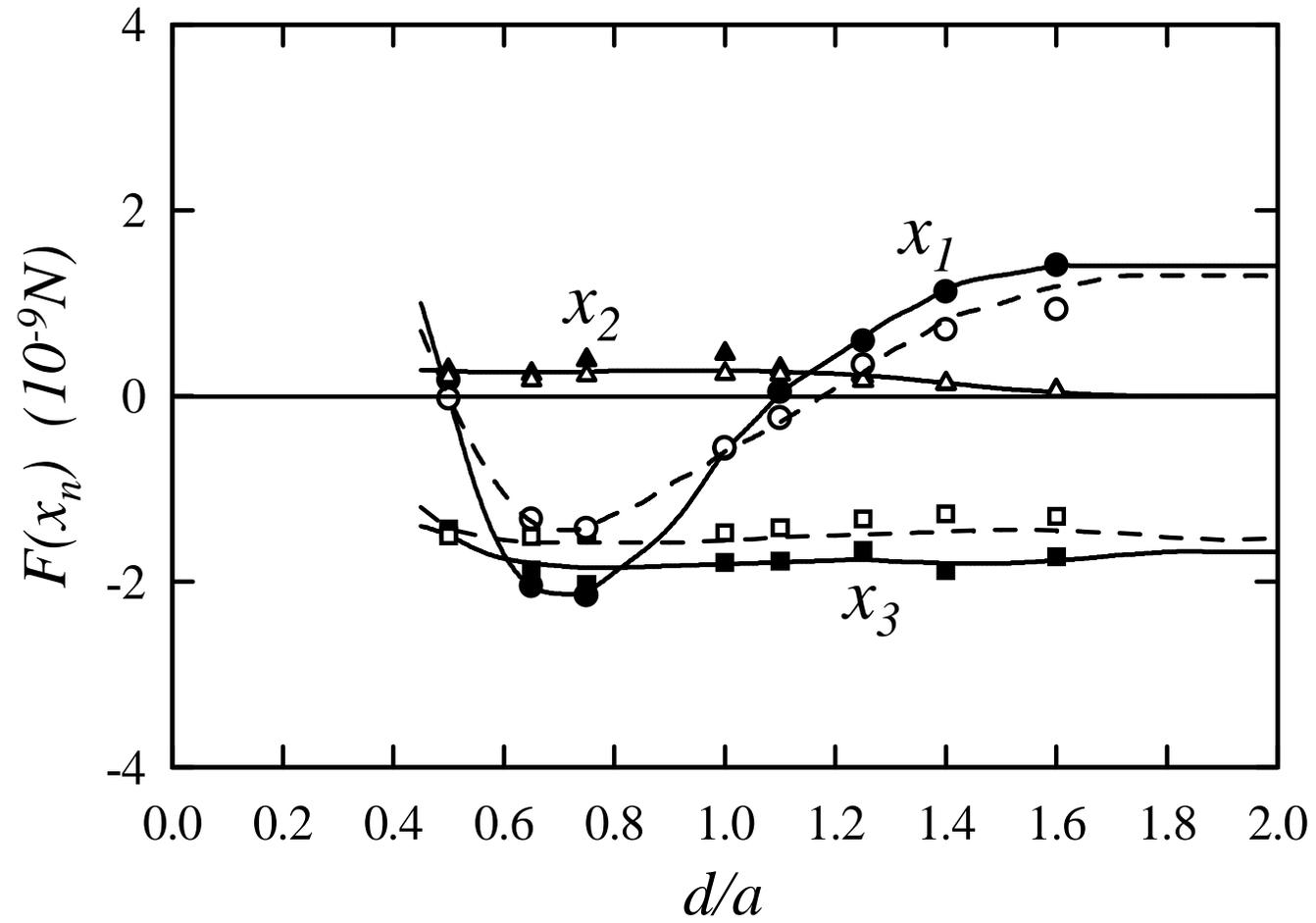
\begin{figure}
\caption{
Atomic forces, $F(x_n)$, acting on the atom
of the $x_n$ layer as a function of the film-film separation.
Solid and open marks stand for the P 
and the AP configurations, respectively.
}
\label{fig4}
\end{figure}

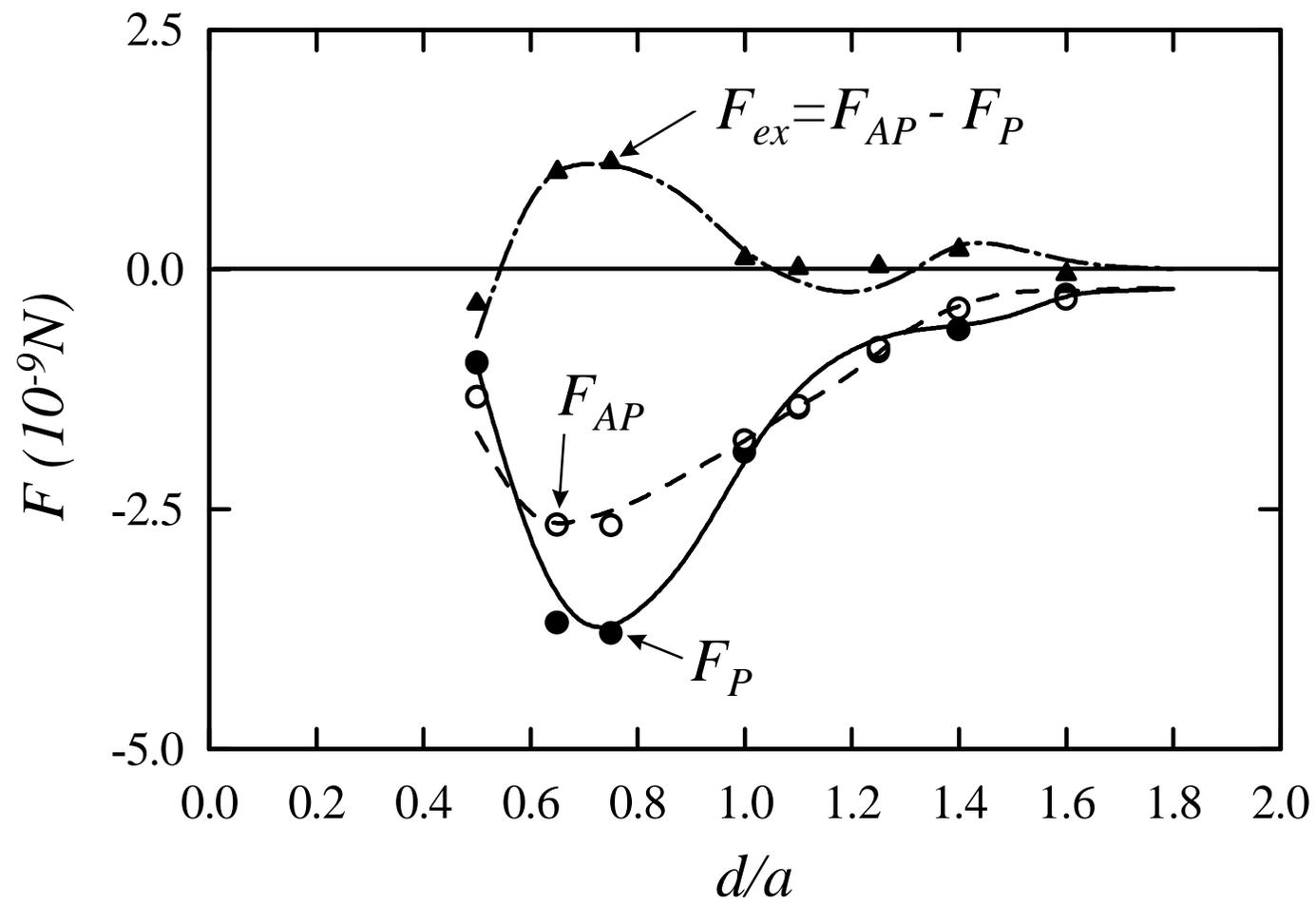
\begin{figure}
\caption{
The film-film separation dependence of
total forces in the P ($F_{\rm P}$: filled circles) 
and AF ($F_{\rm AF}$: open circles)
configurations, and of
the exchange force defined by 
$F_{\rm ex} = F_{\rm AF} - F_{\rm P}$ (triangles).
}
\label{fig5}
\end{figure}

Figure \ref{fig4} shows the atomic forces acting on atoms 
at the $x_1$, $x_2$ and $x_3$ layers of the upper film.
The force direction is perpendicular to the film surface
due to symmetry.
As in the case of magnetic moments (Fig.~\ref{fig2}),
the  atomic force for the $x_1$ layer has
a significant $d$ dependence 
whereas forces for the $x_2$ and $x_3$ layers
are almost independent of $d$.
The total forces summed up these forces:
$F = \sum_n F(x_n)$,
for the P and AF configurations, are shown in Fig.~\ref{fig5}. 
The exchange force defined by $F_{\rm ex} = F_{\rm AP} - F_{\rm P}$
has an oscillation 
against $d/a$ with the period of about $0.7$,
which is identical with the period of the exchange interaction
(Fig.~\ref{fig3}).
The magnitudes of the first and second peaks are $1 \times 10^{-9}$ N
and  $2 \times 10^{-10}$ N, respectively.
In the case of $d/a < 1$, 
where the relevant $d$ orbitals must have a large overlap,
the exchange force may arise from the direct  exchange
couplings between $d$ states of the two films.
In the case of $d/a > 1$, on the other hand,
the exchange coupling is expected to be mediated 
through delocalized $s$ and $p$ electrons.

\begin{center}
{\bf IV. CONCLUSION AND DISCUSSION}
\end{center}

In summary, we have performed the first-principles calculation
of the exchange interaction  and the exchange force
between two magnetic Fe(100) thin films, to show that
the magnitude of the calculated exchange force is  of the 
order of $10^{-9}$ at $d/a < 1$ while
it is of the order of  $10^{-10}$N at $1 < d/a < 1.5$.
It has been shown that
at $d > a$, the perturbation of the 
approaching ferromagnetic film
to magnetic moments in the other film, is negligibly small. 
The calculated magnitude of the
exchange force is in fairly good agreement
with our previous result of $10^{-10} \sim 10^{-11} N$
in spite of a crudeness of the one-dimensional 
electron-gas model.\cite{94Muk}\cite{95Muk}
Our result is also comparable with the exchange force of
$0.4 \times 10^{-9} N$ between a Fe tip and Cr(100) surface
calculated by Ness and Gautier\cite{95Nes}
with the use of the tight-binding model.
Because the sensitivity of the current AFM 
is of the order of $10^{-12}$ to $10^{-13}$N\cite{87Mar2}
our calculation suggests the possibility of the EFM,
although it would be necessary to take into account 
an actual shape of the apex tip before we deduce definite
conclusions. 
Ness and Gautier have been pointed out\cite{95Nes} 
that magnitudes of
the exchange energy and force do not strongly depend on 
the tip morphology, although tip's shape varies
the tip-sample distance  where exchange
interaction and force have  the maximum values.
It has been also shown\cite{95Nes} 
from calculations adopting sharp and blunt tips
that an atomically sharp tip is not necessary for an actual
observation with a constant-force mode,
although a sharp tip is indispensable for a high, lateral 
spatial resolution.
These results\cite{95Nes} justify, 
to some extent, our calculation employing a thin film as a tip;
a first-principles calculation adopting more realistic tips
is under consideration.

In actual measurements of the surface magnetic
structure by using the EFM, we would have the
two possibilities: a constant-force mode and 
a constant-distance mode,
just as in the STM.\cite{82Bin}
In the constant-force mode, we detect the variation of the tip-sample
distance due to the force difference
between the  parallel and anti-parallel configurations.
Such a experiment\cite{90Wie} has been made to detect the
antiferromagnetic  structure from one terrace to the next 
on Cr(100) surface, by using the  constant-current mode in
the SP-STM.
On the contrary, in the constant-distance mode, we directly detect
a change of the  exchange forces between the 
parallel and anti-parallel configurations 
at a given tip-sample separation.
If such a measurement is performed by changing the tip-sample
distance, we would face technical difficulties such as the
snap-in effect: when the tip-sample distance 
comes in the point of force instability
where the gradient of the force 
exceeds the magnitude of the lever stiffness,
the tip and surface would snap into contact.\cite{92Mey}
In order to measure the RKKY-type exchange force, 
we require an exploration of the way to avoid the snap-in effect.
Recently, a continuous measurement of the force curve
between tip and sample,
as approaching towards contact without the snap-in effects,
can be provided by means of 
force-controlled feedback system.\cite{96Jar}
This technique may possibly provide the basis of the EFM measurement.

\section*{Acknowledgements}
This computation in this work has been done
using the facilities of the Supercomputer Center,
Institute for Solid State Physics, University of Tokyo.
This work has also been supported partially by
a Grant-in-Aid for Scientific Researach (B) 
from the Ministry of Education, Science, Sports and Culture of Japan.

\end{document}